\documentclass[a4paper,12pt]{article}
\usepackage{amsfonts}
\usepackage{amsmath}
\numberwithin{equation}{section}

\begin{document}

\title{Surface waves in deformed Bell materials}
\author{Michel Destrade} 
\date{2003}
\maketitle

\bigskip

%
\begin{abstract}

Small amplitude inhomogeneous plane waves are studied as they
propagate on the free surface of a predeformed semi-infinite body
made of Bell constrained material.
The predeformation corresponds to a finite static pure homogeneous
strain. 
The surface wave propagates in a principal direction of strain 
and is attenuated in another principal direction, 
orthogonal to the free surface.
For these waves, the secular equation giving the speed of propagation
is established by the method of first integrals.
This equation is not the same as the secular equation for 
incompressible half-spaces, 
even though the Bell constraint and the incompressibility constraint
coincide in the isotropic infinitesimal limit.

\end{abstract}

\newpage


\section{Introduction}

The so-called `Bell constraint' was established empirically by James
F.~Bell in the 1980s, by conducting a great body of experiments on
certain metals \cite{Bell85}. 
Having subjected  variety of annealed metals to
different finite deformations, he concluded that the condition 
$\text{tr } \mathbf{V} = 3$, where $\mathbf{V}$ is the left
Cauchy--Green tensor, was always satisfied for his samples. 
Later, Beatty and Hayes \cite{BeHa92a}  viewed this constraint 
as a purely kinematical one and used the theory of constrained finite
elasticity to model the behaviour of the materials studied by Bell; 
it turned out that their theoretical predictions agreed remarkably with
the experimental results (a detailed account of these advances, as well
as references to Bell's original articles, can be found in a 
recent review article by Beatty \cite{Beat01}.) 
Hence a new branch of constrained
finite elasticity was born and a substantial literature followed, on
topics such as: finite homogeneous \cite{BeHa92a} and nonhomogeneous
\cite{BeHa92b} deformations, small deformations superimposed on large
\cite{BeHa95a}, universal relations \cite{PuSa96} and motions
\cite{PuSa99}, nonlinear fracture mechanics \cite{Tara97}, stability
analyses \cite{PaBe97,BePa98,PaBe99a,PaBe99b}, Cauchy stress field
analysis \cite{HaSa00}, etc. Small amplitude waves propagating in a
finitely deformed `Bell material' were examined in an unbounded media
for the case of homogeneous plane waves \cite{BeHa95a,PaBe99a} and in
a long circular cylinder for the case of torsional waves
\cite{BeHa95b}. 
The object of this paper is to establish the equation
giving the speed of waves propagating on the plane free surface of a
homogeneously deformed half-space made of Bell material (the secular
equation).

Elastic surface waves on solid half-spaces were first studied by
Rayleigh \cite{Rayl85} in the context of linear isotropic elasticity.
Subsequently, the theory of elastic surface waves evolved in a parallel
manner in the fields of linear anisotropic elasticity 
\cite{Stro58,Stro62} and of finite elasticity under initial stress 
\cite{Biot40,HaRi61}.
Interestingly, although Rayleigh  treated the case of
incompressibility in his paper, surface waves propagating in internally
constrained, triaxially prestressed, elastic materials received 
only tardy attention 
(see for instance Dowhaik and Ogden \cite{DoOg90}, 
Chadwick \cite{Chad97}, or Rogerson \cite{Roge98}.) 
Recently this paper's author proposed a method, 
inspired by Mozhaev \cite{Mozh94}, to derive the explicit
secular equation for surface acoustic waves in monoclinic elastic
crystals \cite{Dest01}. 
In that paper, it was claimed that taking the
first integrals of the tractions rather than of the displacements (as
in \cite{Mozh94}) to obtain the secular equation was a procedure that
could `easily accommodate internal constraints.' 
Here this claim is validated by the consideration of the Bell 
constraint for small amplitude surface waves travelling in a 
principal direction of a homogeneously deformed half-space. 

In section 2, the incremental equations corresponding to a 
small amplitude displacement superposed on a large deformation 
\cite{BeHa95a} are recalled.
These equations are specialized to the consideration of a
surface wave in a deformed Bell material in section 3,
where the equations of motion and the boundary conditions 
are written for the tractions on planes parallel to the free surface.
Next in section 4, the secular equation is obtained from the 
vanishing of a $ 2 \times 2$ certain determinant. 
This secular equation is compared to that obtained for incompressible
\textit{finitely} prestressed bodies \cite{DoOg90} 
and it is found that one secular equation may not be deduced 
from the other.
However, because a Bell material behaves in \textit{infinitesimal} 
motions like an incompressible one \cite{BeHa92a}, the
secular equation written when the half-space is undeformed should
coincide with that found by Rayleigh for incompressible isotropic
materials -- it is proved that such is indeed the case.

\section{Small motion superposed on a large deformation}

Beatty and Hayes established the equations governing the elastic 
behaviour of a Bell constrained material subjected to finite 
deformations \cite{BeHa92a,BeHa92b}, as well as those 
describing small deformations superimposed on large \cite{BeHa95b}.
Here their results are recalled in the context of small amplitude
plane waves propagating in a homogeneously deformed Bell constrained
half-space.

For a general hyperelastic Bell material, the strain energy density
$\Sigma$ depends only on $I_2$ and $I_3$, the respective second and
third invariants of the left Cauchy--Green tensor $\mathbf{V}$,
\begin{equation} \label{Sigma}
\Sigma = \Sigma(I_2,I_3), \text{ where } 
I_2 = [ (\text{tr } \mathbf{V})^2 - \text{tr } (\mathbf{V}^2)]/2, \:
I_3 = \text{det } \mathbf{V}.
\end{equation}
Also, the Bell constraint imposes that $I_1$, the first invariant of
$\mathbf{V}$, satisfy
\begin{equation} \label{Bell}
I_1 = \text{tr } \mathbf{V} = 3,
\end{equation}
for all deformations and at all times.
Because of this constraint, a workless stress $p \mathbf{V}$ 
is introduced in the stress-strain relationship, where the scalar 
$p$ is to be determined from the equations of motion and boundary and 
initial conditions. 
Hence the constitutive equation relates the Cauchy stress tensor
$\mathbf{T}$ to the left Cauchy--Green tensor as \cite{BeHa92a}
\begin{equation} \label{constitutive}
 \mathbf{T}= p \mathbf{V} + \omega_0 \mathbf{1} 
  +  \omega_2 \mathbf{V}^2. 
\end{equation}
Here, the material response functions $ \omega_0$ and $ \omega_2$
are defined by
\begin{equation} \label{omega}
\omega_0 = \frac{\partial \Sigma}{\partial I_3},
\quad 
\omega_2 = - \frac{1}{I_3} \frac{\partial \Sigma}{\partial I_2},
\end{equation}
and these quantities should satisfy the Beatty--Hayes 
$A$-inequalities \cite{BeHa92a}
\begin{equation} \label{A-inequalities}
\omega_0 (I_2,I_3) \le 0, \quad \omega_2 (I_2,I_3) > 0.
\end{equation}

A solid half-space made of hyperelastic Bell material is now 
considered.
In an isotropic undeformed reference state, the positions of its 
material particles are denoted by $\mathbf{X}$.
Next, the half-space is subjected to a finite static pure homogeneous 
deformation, so that particles originally at $\mathbf{X}$ have
moved to $\mathbf{x}= \mathbf{x}(\mathbf{X},t)$.
Using the principal directions of strain as a rectangular Cartesian 
coordinate system, this deformation is expressed as 
$x_{\alpha} = \lambda_{\alpha} X_{\alpha}$ ($\alpha = 1,2,3$, no sum),
where the principal stretch ratios $\lambda_1, \lambda_2, \lambda_3$
satisfy $\lambda_1+ \lambda_2+ \lambda_3 = 3$, according to 
\eqref{Bell}. 
Also, for this deformation, 
\begin{equation} \label{invariants-static}
I_2 = \lambda_1 \lambda_2+ \lambda_2 \lambda_3+ \lambda_3 \lambda_1,
\quad
I_3 = \lambda_1 \lambda_2 \lambda_3.
\end{equation}

The surface $x_2 = 0$ is assumed to be free of tractions and the 
following Cauchy stress tensor $\mathbf{T}_o$ satisfies the equations
of equilibrium \cite{BeHa92a},
\begin{equation}  \label{T}
\begin{array}{l}
(T_{o})_{\alpha \alpha} = p_o \lambda_{\alpha} + \omega_0  
  + \lambda_{\alpha}^2 \omega_2, \quad  \alpha = 1,2,3, \\
(T_{o})_{\alpha \beta} = 0, \quad  \alpha \ne \beta,
\end{array}
\end{equation}
where $ \omega_0$ and $ \omega_2$ are evaluated at $I_2$, $I_3$ 
given by \eqref{invariants-static}, and 
\begin{equation} \label{po}
p_o = - ( \omega_0   + \lambda_2^2 \omega_2)/ \lambda_2.
\end{equation}

Next, the particle at $\mathbf{x}$ undergoes an incremental 
displacement $\epsilon \mathbf{u}(\mathbf{x},t)$
(with associated incremental constraint parameter 
$\epsilon p^*(\mathbf{x},t)$), so that this further motion is
of the form $\mathbf{x}+ \epsilon \mathbf{u}(\mathbf{x},t)$
(and the constraint parameter is of the form
$p_o + \epsilon p^*(\mathbf{x},t)$), where $\epsilon$ is a small 
parameter. 
Throughout the rest of the paper, terms of order higher 
than $\epsilon$ are neglected.
The corresponding incremental Cauchy stress $\mathbf{T}^*$ is given
by \cite{BeHa95a}:
\begin{align} \label{T*}
& T^*_{11} = \lambda_1 p^* 
  + ( \lambda_1 p_o + C_{11}) \frac{\partial u_1}{\partial x_1} 
  + C_{12} \frac{\partial u_2}{\partial x_2} 
  + C_{13} \frac{\partial u_3}{\partial x_3},  \nonumber \\
& T^*_{22} = \lambda_2 p^* 
  + C_{21} \frac{\partial u_1}{\partial x_1} 
  + ( \lambda_2 p_o + C_{22}) \frac{\partial u_2}{\partial x_2} 
  + C_{23} \frac{\partial u_3}{\partial x_3}, \nonumber  \\ 
& T^*_{33} = \lambda_3 p^* 
  + C_{31} \frac{\partial u_1}{\partial x_1} 
  + C_{32} \frac{\partial u_2}{\partial x_2}
  + ( \lambda_3 p_o + C_{33}) \frac{\partial u_3}{\partial x_3},  \\
& T^*_{12} =
   b_3 (\lambda_2^2 \frac{\partial u_1}{\partial x_2} 
           + \lambda_1^2 \frac{\partial u_2}{\partial x_1}),
 \nonumber \\ 
& T^*_{23} =
   b_1 (\lambda_3^2 \frac{\partial u_2}{\partial x_3}
           + \lambda_2^2 \frac{\partial u_3}{\partial x_2} ), 
\nonumber \\
& T^*_{13} =
   b_2 (\lambda_3^2 \frac{\partial u_1}{\partial x_3}
           + \lambda_1^2 \frac{\partial u_3}{\partial x_1}), \nonumber 
\end{align} 
with
\begin{equation}	\label{C-b}
\begin{array}{l}
C_{\alpha \beta} = 2 \lambda^2_{\alpha} \delta_{\alpha \beta} \omega_2
    - \lambda_{\beta}^2 (\omega_{02} + \lambda_{\alpha}^2 \omega_{22})
    + \lambda_1 \lambda_2  \lambda_3 (\omega_{03} + 
                                     \lambda_{\alpha}^2 \omega_{23}),
\\
b_{\gamma} = \omega_2 + p_o/(3-\lambda_{\gamma}),
\end{array}
\end{equation}
where the derivatives $\omega_{0\Gamma}$, $\omega_{2\Gamma}$ 
($\Gamma = 2,3$) of the material response functions $\omega_0$,
$\omega_2$ are taken with respect to $I_\Gamma$ and evaluated at 
$I_2$, $I_3$ given by \eqref{invariants-static}.
The incremental Bell constraint yields
\begin{equation}	\label{incrBell}
\lambda_1 \frac{\partial u_1}{\partial x_1} 
  + \lambda_2 \frac{\partial u_2}{\partial x_2} 
      + \lambda_3 \frac{\partial u_3}{\partial x_3} = 0;
\end{equation}
and the components of the incremental traction 
on planes parallel to the free surface $x_2=0$ are \cite{BeHa95a}
\begin{align}	\label{tractions}
& t^*_{12}= T^*_{12} 
           - (\lambda_1 p_o  + \omega_0 + \lambda_1^2 \omega_2) 
                \frac{\partial u_2}{\partial x_1} , \nonumber \\
& t^*_{22}= T^*_{22},  \\
& t^*_{32}= T^*_{32} 
           - (\lambda_3 p_o + \omega_0 + \lambda_3^2 \omega_2) 
                \frac{\partial u_3}{\partial x_2} . \nonumber
\end{align}

Finally, the incremental equations of motion read
\begin{equation}	\label{incrMotion}
 \frac{\partial T^*_{ij}}{\partial x_j}
   = \rho \frac{\partial^2 u_i}{\partial t^2},
\end{equation}
where $\rho$ is the material density of the half-space, when it is
maintained in the state of static homogeneous deformation. 

\section{Principal surface waves}

Now the analysis of small amplitude motions in a deformed Bell 
half-space is specialized to the consideration of a  
plane wave propagating sinusoidally on the free surface
$x_2=0$ in the $x_1$--direction, with attenuation in the 
 $x_2$--direction.
Without loss of generality, the component $u_3$ is taken to be zero.

Hence the wave may be modeled as 
\begin{equation} 
u_{\Gamma}(x_1,x_2,t) = U_{\Gamma}(x_2)e^{ik(x_1-vt)}, 
\quad 
p^*(x_1,x_2,t) = k	P(x_2) e^{ik(x_1-vt)},
\end{equation}
where $\Gamma = 1,2$, $k$ is the real wave number, 
$v$ is the real wave speed of propagation, 
and the specific dependence of the amplitudes $U_{\Gamma}$
and $P$ on $x_2$ need not be specified \cite{Mozh94}.
Consequently, the incremental Bell constraint \eqref{incrBell} 
now reduces to
\begin{equation}	\label{Bell2}
i \lambda_1 U_1 + \lambda_2 U_2' = 0,
\end{equation}
where, here and henceforward, the prime denotes differentiation 
with respect to $kx_2$. 
The other incremental expressions recalled in the previous section 
are also simplified.

The incremental Cauchy stress components are now given by
\begin{align} \label{T*surface}
& T^*_{11} = k[\lambda_1 P 
                + i( \lambda_1 p_o + C_{11}) U_1 +  C_{12} U_2'],
  \nonumber \\
& T^*_{22} =  k[\lambda_2 P 
                + iC_{21} U_1 +  ( \lambda_2 p_o + C_{22}) U_2'],
  \nonumber \\
& T^*_{33} =  k[\lambda_3 P 
                + iC_{31} U_1 + C_{32}) U_2'],
   \\
& T^*_{12} = k b_3 (\lambda_2^2 U_1' + i \lambda_1^2 U_2),
 \nonumber \\ 
& T^*_{23} =   T^*_{13} = 0. \nonumber 
\end{align} 
The third incremental equation of motion \eqref{incrMotion}$_3$ is
automatically satisfied, while the two others reduce to 
\begin{equation} \label{Motion2}
 i T^*_{1\Gamma} + (T^*_{\Gamma 2})' = - k \rho v^2 U_\Gamma, \quad
(\Gamma = 1,2).
\end{equation}

Also, the expressions for the tractions 
$t^*_{12}$, $t^*_{22}$, $t^*_{23}$ \eqref{tractions} have reduced to 
\begin{equation}	\label{tractions2}
t^*_{12} = T^*_{12} 
           - ik ( \lambda_1 p_o + \omega_0 + \lambda_1^2 \omega_2)U_2,
\quad
t^*_{22} = T^*_{22},
\quad
t^*_{23} = T^*_{23} = 0.
\end{equation}

Now a system of four first order differential equations can be written,
for the nonzero displacements $U_1$, $U_2$, and for the quantities 
$t_1$, $t_2$, defined in terms of the tractions as
\begin{equation} 
t_1 = k^{-1} t^*_{12}, \quad t_2 = k^{-1} t^*_{22}.
\end{equation}

Taking equations \eqref{po}, \eqref{C-b}, and \eqref{Bell2} 
into account, this system can be written as
\begin{align} \label{quartic}
& t_1' + i \lambda_1 \lambda_2^{-1} t_2  
 - (\lambda_1 \lambda_2^{-1}C - \rho v^2) U_1 = 0,  
\nonumber \\
& t_2' + i  t_1 
 	-[b_3(\lambda_1^2 -  \lambda_2^2) - \rho v^2 ] U_2= 0, 
\\
& U_2' + i \lambda_1 \lambda_2^{-1} U_1 = 0,  \nonumber \\
& b_3 \lambda_2^2 U_1' + ib_3 \lambda_2^2 U_2 -  t_1 = 0, 
\nonumber
\end{align}
where  $C$ is defined by 
\begin{equation} \label{C}
C = \lambda_1^{-1} \lambda_2 C_{11} 
	+ \lambda_1 \lambda_2^{-1} C_{22} -C_{12} - C_{21}
		-2\omega_0 - (\lambda_1^2+\lambda_2^2)\omega_2,
\end{equation}
and the following expression for $b_3 \lambda_2^2$, obtained
from \eqref{C-b}, and \eqref{po}, has also been used,
\begin{equation} \label{b3}
b_3 \lambda_2^2 = 
    - (\omega_0 - 
	\lambda_1 \lambda_2 \omega_2)/(1 +\lambda_1 \lambda_2^{-1}).
\end{equation}

The system must be solved when subject to the following 
boundary conditions.
First, the surface $x_2=0$ remains free of traction, so that
\begin{equation}	\label{BC1}
t^*_{\Gamma 2} (0) = 0, \quad (\Gamma = 1,2);
\end{equation}
and second, the displacement must vanish at infinite distance from the 
free surface, so that 
\begin{equation}	\label{BC2}
U_{\Gamma} (\infty) = 0, \quad P(\infty) = 0, \quad (\Gamma = 1,2).
\end{equation}

\section{Secular equation}

Here the secular equation is derived by applying the method of first
integrals to the system \eqref{quartic}.

Proceed as follows.
First, rewrite the boundary conditions \eqref{BC1}-\eqref{BC2} as
\begin{equation} \label{BC3}
t_\Gamma(0) = t_\Gamma(\infty) = 0, \quad (\Gamma = 1,2).
\end{equation}
These last conditions, together with \eqref{quartic}$_{1,2}$, 
imply that
\begin{equation} \label{BC4}
t'_\Gamma(\infty) = 0, \quad (\Gamma = 1,2).
\end{equation} 
Next, obtain $U_\Gamma$  in terms of $t_\Gamma$, $t'_\Gamma$
from \eqref{quartic}$_{1,2}$, and $U'_\Gamma$  in terms of $U_\Gamma$, 
$t_\Gamma$ from \eqref{quartic}$_{3,4}$, or equivalently, 
in terms of $t_\Gamma$, $t'_\Gamma$.
Finally, substitute these expressions into the derivatives of 
equations \eqref{quartic}$_{1,2}$ to get,
\begin{align} \label{2ndOrder}
& [b_3(\lambda_1^2-\lambda_2^2) - \rho v^2] t_1''
  + i \beta_{12} t_2'
    -\frac{1}{b_3\lambda_2^2}
      (\lambda_1 \lambda_2^{-1}C -\rho v^2)
	(b_3\lambda_1^2 -\rho v^2) t_1 = 0, \\
&  (\lambda_1\lambda_2^{-1}C -\rho v^2) t_2''
  + i \beta_{12} t_1'
    -\lambda_1^2\lambda_2^{-2}
	[b_3(\lambda_1^2-\lambda_2^2) - \rho v^2] t_2 = 0, \nonumber
\end{align} 
where \eqref{b3} has been used, and $\beta_{12}$ is defined by
\begin{equation}
\beta_{12} = \lambda_1\lambda_2^{-1} [b_3(\lambda_1^2-\lambda_2^2)+C]
	- (1+\lambda_1\lambda_2^{-1})\rho v^2.
\end{equation}

The quantity $\beta_{12}$ does not play any role in the secular 
equation, as is now seen.
Multiply \eqref{2ndOrder}$_1$ by $t_1'$ and \eqref{2ndOrder}$_2$ 
by $t_2'$, and integrate with respect to $k x_2$ 
between $x_2=0$ and $x_2=\infty$, to obtain, using \eqref{BC3} 
and \eqref{BC4},
\begin{multline}
 [b_3(\lambda_1^2-\lambda_2^2) - \rho v^2] t_1'(0)^2
  - 2 i \beta_{12} \textstyle{\int} t_1' t_2' = 0, \\
  (\lambda_1\lambda_2^{-1}C -\rho v^2) t_2'(0)^2
  - 2i \beta_{12} \textstyle{\int} t_1' t_2' = 0, 
\end{multline}
and therefore,
\begin{equation}  \label{eq1}
[b_3(\lambda_1^2-\lambda_2^2) - \rho v^2] t_1'(0)^2
 - (\lambda_1\lambda_2^{-1}C -\rho v^2) t_2'(0)^2 = 0.
\end{equation}
Similarly, multiply \eqref{2ndOrder}$_1$ by 
$[b_3(\lambda_1^2-\lambda_2^2) - \rho v^2]t_1'+ i \beta_{12} t_2$ 
and \eqref{2ndOrder}$_2$ 
by $(\lambda_1\lambda_2^{-1}C -\rho v^2) t_2' + i \beta_{12} t_1$, 
and integrate with respect to $k x_2$ 
between $x_2=0$ and $x_2=\infty$, to obtain, using \eqref{BC3} 
and \eqref{BC4},
\begin{multline}
 [b_3(\lambda_1^2-\lambda_2^2) - \rho v^2]^2 t_1'(0)^2
  - 2 i \beta_{12}\frac{1}{b_3\lambda_2^2}
      (\lambda_1 \lambda_2^{-1}C -\rho v^2)
	(b_3\lambda_1^2 -\rho v^2) \textstyle{\int} t_1 t_2 = 0, \\
  (\lambda_1\lambda_2^{-1}C -\rho v^2)^2 t_2'(0)^2
  - 2i \beta_{12}\lambda_1^2\lambda_2^{-2}
	[b_3(\lambda_1^2-\lambda_2^2) - \rho v^2] 
		\textstyle{\int} t_1 t_2 = 0, 
\end{multline}
and therefore,
\begin{multline}  \label{eq2}
[b_3(\lambda_1^2-\lambda_2^2) - \rho v^2]^3 
	\lambda_1^2\lambda_2^{-2} t_1'(0)^2 \\
 - \frac{1}{b_3\lambda_2^2}(\lambda_1\lambda_2^{-1}C -\rho v^2)^3
	(b_3\lambda_1^2 -\rho v^2)t_2'(0)^2 = 0.
\end{multline}

The two homogeneous equations \eqref{eq1} and \eqref{eq2} for the two
unknowns $t_1'(0)^2$ and $t_2'(0)^2$ yield nontrivial solutions 
only when the corresponding determinant is zero.
This condition provides the \textit{secular equation for surface
waves propagating in a principal direction of a deformed Bell
material} as,
\begin{equation} \label{secular}
(b_3 \lambda_1^2 - \rho v^2) (\lambda_1\lambda_2^{-1}C -\rho v^2)^2
 = b_3 \lambda_1^2[b_3 (\lambda_1^2 - \lambda_2^2) - \rho v^2]^2.
\end{equation}

This equation is distinct from that obtained for deformed 
\textit{incompressible} materials by Dowaikh and Ogden \cite{DoOg90}, 
which reads as
\begin{equation} \label{incompressible}
(\widetilde{b}_3 \lambda_1^2 - \rho v^2) (\widetilde{C} -\rho v^2)^2
 = \widetilde{b}_3 \lambda_2^2
	[\widetilde{b}_3 (\lambda_1^2 - \lambda_2^2) - \rho v^2]^2,
\end{equation}
where $\widetilde{b}_3$ and $\widetilde{C}$ are defined in terms of
the stretch ratios $\lambda_1$, $\lambda_2$ and the stored energy 
density $W$ of the incompressible material as
\begin{align}
& \widetilde{b}_3 =
 (\lambda_1 \frac{\partial W}{\partial \lambda_1} - 
	\lambda_2 \frac{\partial W}{\partial \lambda_2})
 		/(\lambda_1^2 - \lambda_2^2), 
\nonumber \\
& \widetilde{C} =
 \lambda_1^2 \frac{\partial^2 W}{\partial \lambda_1^2} - 
 2\lambda_1\lambda_2 
	\frac{\partial^2 W}{\partial \lambda_1\partial \lambda_2} +
 \lambda_2^2 \frac{\partial^2 W}{\partial \lambda_2^2} +
 2\lambda_2 \frac{\partial W}{\partial \lambda_2}
\end{align}
Although equations \eqref{secular} and \eqref{incompressible}
share a similar left hand-side, their right hand-sides are different
and could not have been a priori deduced one from another.
In particular, $b_3$ and $\lambda_1\lambda_2^{-1}C$ can formally
be transformed into $\widetilde{b}_3$ and $\widetilde{C}$ respectively
to get from the left hand-side of \eqref{secular} to the 
left hand-side of \eqref{incompressible}, but this transformation
does not work for the right hand-sides of the equations, 
because the factor $\lambda_1^2$ is replaced by $\lambda_2^2$.
Hence, when the influence of the prestrain on the propagation of 
surface waves is studied \cite{DoOg90}, the behaviour of Bell
materials is in contrast to that of incompressible materials,
except in the special degenerate case of a bi-axial prestrain such
that $\lambda_1 = \lambda_2$.

However, when the Bell material is \textit{unstrained} for the problem
at hand ($\lambda_1 = \lambda_2 = \lambda_3 = 1$),  
only infinitesimal deformations (the surface waves) occur in the 
now isotropic half-space.
In such a case, Beatty and Hayes \cite{BeHa92a} show
that the Bell constraint is equivalent to the incompressibility
constraint. 
Here, the quantities $C$, $b_3$ are then given by
\begin{equation} 
\lambda_1\lambda_2^{-1}C = C = 2(\omega_2 - \omega_0) = 4 b_3,
\end{equation}
and the secular equation \eqref{secular} reduces to 
\begin{equation} 
(1 - \xi)(4 - \xi)^2 = \xi^2 , \quad
	\text{or } \quad
\xi^3 - 8\xi^2 + 24\xi - 16 = 0,
\end{equation}
where $\xi = 2 \rho v^2 / (\omega_2 - \omega_0)$.
This last equation is that established by Rayleigh \cite{Rayl85}
for surface waves in isotropic linearly elastic incompressible 
half-spaces.


\end{document}